\documentclass[aps,pre,showpacs,twocolumn,superscriptaddress]{revtex4}
\usepackage[dvips]{graphicx}

\begin{document}

\title{Effects of Transport Memory and Nonlinear Damping in a \\
Generalized Fisher's Equation}

\date{\today}

\author{G. Abramson}
\altaffiliation[Permanent Address: ]{Centro At{\'o}mico Bariloche, Instituto
Balseiro and CONICET, 8400 S. C. de Bariloche, Argentina}
\email{abramson@cab.cnea.gov.ar}
\affiliation{Center for Advanced Studies and Department of Physics
and Astronomy, University of New Mexico, Albuquerque, New Mexico 87131}

\author{A. R. Bishop}
\affiliation{Theoretical Division and Center for Nonlinear Studies,
Los Alamos National Laboratory, Los Alamos, New Mexico 87545}

\author{V. M. Kenkre}
\affiliation{Center for Advanced Studies and Department of Physics
and Astronomy, University of New Mexico, Albuquerque, New Mexico 87131}

\begin{abstract}
Memory effects in transport require, for their incorporation into reaction
diffusion investigations, a generalization of traditional equations. The
well-known Fisher's equation, which combines diffusion with a logistic
nonlinearity, is generalized to include memory effects and traveling wave
solutions of the equation are found. Comparison is made with alternate
generalization procedures.
\end{abstract}


\maketitle

\section{Introduction}

Fisher's equation \cite{murray} describes the dynamics of a field $u(x,t)$
subject to diffusive transport and logistic growth:
\begin{equation}
\frac{\partial u}{\partial t}= D \frac{\partial^2 u}{\partial x^2}+ k u(1-u/K).
\label{fisher}
\end{equation}
This kind of reaction-diffusion equation is relevant in chemical kinetics as
well as in ecological contexts where $u$ diffuses via a diffusion constant $D$
and grows with a linear growth rate $k$ while the environment imposes a
carrying capacity $K$. Equation~(\ref{fisher}) belongs to a family of single
component models of broad applicability. Fisher proposed it as a deterministic
model of the spread of a favored gene in a population~\cite{fisher36}. Other
systems in which Eq.~(\ref{fisher}) plays a significant role include flame
propagation, neutron flux in a nuclear reactor and the dynamics of defects in
nematic liquid crystals~\cite{murray,mikhailov,canosa,saarloos,borzi}. Work has
been done on the propagation of wave-like structures in unbounded systems
\cite{borzi,hassan}, as well as the formation of spatial structures in finite
geometries \cite{izus}.

Diffusion is typically a limit of a more coherent motion interrupted by
scattering events which is valid when the scattering events are extremely
frequent. A useful manner of describing situations in which the frequency of
scattering events is not infinitely large is through the incorporation of
memories with finite decay times \cite{kenkrereineker}. An analytical
investigation of reaction-diffusion phenomena based on the introduction of
exponential memory in the context of the logistic nonlinearity of the Fisher's
equation, and a piecewise linearity approximation, was recently given by Manne,
Hurd and Kenkre (MHK) \cite{MHK}. A number of interesting results appeared in
that analysis, including a generalization of the relation between the minimum
speed of shock fronts and the system parameters, and the appearance of a speed
limit beyond which the fronts would exhibit spatial oscillations.

As stated in Ref.~\cite{MHK}, the generalization of Fisher's equation is not
unique, the one employed in Ref.~\cite{MHK} having been chosen for analytic
tractability. A more natural generalization results in a more complex structure
of the equation, involves a nonlinearity in the damping, and is the subject of
the present paper. Our investigations show that, in the light of the present
generalization, some of the interesting predictions of Ref.~\cite{MHK} are
modified drastically, while others are only slightly changed.

Reference~\cite{MHK} first generalized the linear part of Eq.~(\ref{fisher}) by
incorporating an exponential memory, converted the resulting
integro-differential equation via differentiation into a differential
(specifically the telegrapher's) equation, and finally added the nonlinear
logistic term to obtain the starting point of the analysis. Here, however, we
take as our point of departure,
\begin{equation}
\frac{\partial u}{\partial t}= D \int_0^t \phi(t-\tau)
\frac{\partial^2 u}{\partial x^2} d\tau+ k f(u),
\label{fisher-memory}
\end{equation}
which reduces to Fisher's equation if the nonlinearity is logistic, and if the
memory function $\phi(t)$ is a $\delta$-function in time. Generally, the decay
time of $\phi(t)$ is a measure of the time between scattering events: we will
take $\phi(t)$ to have  a simple exponential form $\phi(t)=\alpha e^{-\alpha
t}$, where $1/\alpha$ represents the scattering time.

Transformation of Eq.~(\ref{fisher-memory}) into a differential equation is
trivially done by differentiating once with respect to time:
\begin{equation}
\frac{\partial^2 u}{\partial t^2}+
[\alpha-k f'(u)]\,\frac{\partial u}{\partial t} = v^2 \,\frac{\partial^2
u}{\partial x^2} +\alpha k f(u),
\label{telegraph}
\end{equation}
where, as in \cite{MHK}, we  put $D\alpha=v^2$, the physical meaning of $v$
being the speed dictated by the medium in the absence of scattering. This is
the speed at which the underlying quasiparticle, whose number density or
probability density is described by $u(x,t)$, moves ballistically (coherently)
in between scattering events. If the coherent motion is interrupted too often
by scattering, the motion looks diffusive and one returns to the Fisher limit.

Equation (\ref{telegraph}) becomes, in the absence of the nonlinearity, $k=0$,
the well-known telegrapher's equation \cite{morse} suggested by Lord Kelvin for
the description of propagation of transatlantic telegraph signals. Equation
(\ref{telegraph}) differs from the starting point of the MHK analysis, viz.,
\begin{equation}
\frac{\partial^2 u}{\partial t^2}+
\alpha \frac{\partial u}{\partial t} = v^2 \,\frac{\partial^2
u}{\partial x^2} +\alpha k f(u),
\label{mhkteleg}
\end{equation}
in a significant way. The former shows that the damping coefficient multiplying
$\partial u/\partial t$ is not constant, and can even be negative.

Whereas Eq.~(\ref{mhkteleg}) is obviously simpler in form than
Eq.~(\ref{telegraph}), the latter could be argued to  stem from a more natural
generalization of the Fisher equation, involving a simple replacement of a
$\delta$-function memory by a finite decay time memory in the transport term
$D\partial^2u/\partial x^2$ in~(\ref{fisher}). We are interested in comparing
the physical consequences of the two generalizations in the context of
traveling wave solutions. We do this in two steps. In Section~\ref{exact}, we
give a general analysis of traveling wave solutions without approximating the
equation in any way. We investigate the minimum speed of wavefronts, the
existence of spatially oscillatory  fronts given in MHK, and give a qualitative
description of all the kinds of front shapes that the system can support as a
function of the parameters. In a certain regime in parameter space, a dynamical
stabilization of the unstable state of the logistic equation is observed, with
fronts of state $u=0$ invading the state $u=K$. In Section~\ref{piecewise}, we
replace the nonlinearity by a piecewise linear approximate form and obtain
analytic solutions following the method of MHK. We end with concluding remarks
in Section~\ref{conclusion}.

\section{Nonlinear analysis}
\label{exact}

We look for waves moving in the direction of increasing $x$: $u(x,t)=K\,
U(x-ct)=K\,U(z)$, where $c$ is the speed of the nonlinear wave, generally
different from the speed of linear waves $v$, dictated by the medium. We are
also renormalizing with respect to the carrying capacity $K$ to simplify the
reaction term in the logistic case. With this ansatz, we obtain from
(\ref{telegraph}) the following ordinary differential equation:
\begin{equation}
(v^2-c^2)\,U'' + c\,[\alpha-kf'(U)]\,U' + \alpha k f(U) =0.
\label{oscillator}
\end{equation}

For a logistic reaction term: $f(U)=U(1-U)$ and $f'(U)=1-2U$, so that
Eq.~(\ref{oscillator}) becomes:
\begin{equation}
m U'' +c\, (\alpha-k+2kU)\,U' +\alpha k\, U(1-U)=0,
\label{oscillator2}
\end{equation}
where we have followed the notation of MHK, $m=v^2-c^2$, to emphasize the
formal similarity between (\ref{oscillator2}) and the equation of motion of a
damped oscillator of mass $m$, subject to the nonlinear force $-\alpha
kU(1-U)$.

A convenient way to analyze the solutions of (\ref{oscillator2}) is to write it
as a first order system:
\begin{equation}
\begin{array}{rll}
U'&=V&\equiv f(U,V)\\
V'&=\frac{1}{m}[(k-\alpha-2kU)c\,V-\alpha k\,U(1-U)]&\equiv g(U,V).
\end{array}
\label{system1}
\end{equation}
The system (\ref{system1}) has two equilibria: $(U^*,V^*)=(0,0)$ and
$(U^*,V^*)=(1,0)$. We can analyze the character of these by looking at the
linear behavior in a neighborhood of the equilibria.

\begin{figure}
\centering
\resizebox{\columnwidth}{!}{\includegraphics[bb=8 200 784 793]{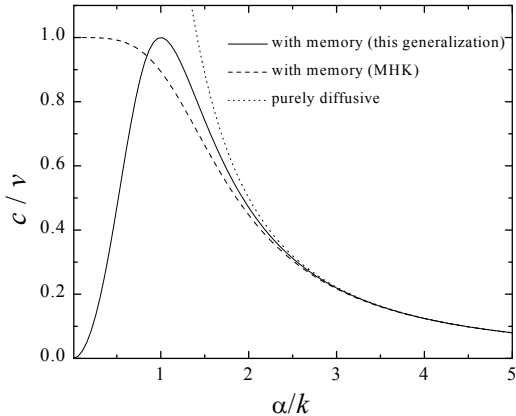}}
\caption{The ratio of the minimum speed of nonlinear waves $c_{min}$ to
the medium-dictated speed $v$, as a function of the system parameter
$\alpha/k$, which is the ratio of the scattering rate at which motion coherence
is interrupted to the growth rate. Lower values of $\alpha/k$ represent more
coherent motion. Three curves are shown: the present generalization (full
line), the MHK generalization of Ref.~\protect\cite{MHK} (dashed line), and the
purely diffusive Fisher limit (dotted line) with $D=v^2/\alpha$.}
\label{cmin-fig}
\end{figure}

At the equilibrium $(0,0)$ we have the eigenvalues:
\begin{equation}
\lambda_\mp=\frac{c(k-\alpha)\mp\sqrt{(k-\alpha)^2c^2-4k\alpha m}}{2m}.
\end{equation}
If we assume that $U$ is a density or a concentration, solutions where $U$
oscillates below $U=0$ are not allowed. This imposes a condition on the
eigenvalues $\lambda_\mp$ to be real, from which we obtain:
\begin{equation}
c\ge c_{min}=v\frac{1}{\sqrt{1+\frac{1}{4}(y-1/y)^2}},
\label{cmin}
\end{equation}
where $y=\sqrt{\alpha/k}$. This relation states that there is a minimum value
of the speed of the nonlinear waves. It is to be compared to the known result
in the context of Fisher's equation: $c_{min}=2\sqrt{kD}$ (see for
example~\cite{murray}). If we make the formal identification $D\alpha=v^2$, as
in (\ref{telegraph}), we have:
\begin{equation}
c\ge c_{min}=2v\sqrt{1/y}.
\end{equation}
Equation~(\ref{cmin}) is also to be compared to the previously found result of
Ref.~\cite{MHK}:
\begin{equation}
c\ge c_{min}^{MHK}=v\frac{1}{\sqrt{1+\frac{1}{4}y^2}},
\label{cminmhk}
\end{equation}
Figure~\ref{cmin-fig} displays a comparison of the minimum speeds $c_{min}$ as
a function of the system parameter $\alpha/k$ in the three cases: purely
diffusive (Fisher limit), the MHK generalization, and the present
generalization. It can be seen that the two generalizations approach
asymptotically the behavior of the purely diffusive situation for large values
of $\alpha/k$. This is to be expected since $\alpha\rightarrow\infty$,
$v\rightarrow\infty$, $v^2/\alpha=D$ is the diffusive limit of
Eq.~(\ref{fisher-memory})]. At low values of the ratio $\alpha/k$, however,
there is a notable difference. The finite correlation time generalizations
allow waves with lower speed than those present in Fisher's equation. Moreover,
the present generalization predicts a sharply different behavior in the region
$\alpha <k$, due to the anti-damping present in Eq.~(\ref{oscillator2}). This
growing branch of the $c_{min}$ function turns out to have drastic consequences
in the nature of the traveling fronts when the system is far from the diffusive
regime. On the one hand, it can be seen that waves with speed $c=v$ are
obtained at a finite value of the memory constant $\alpha$, namely $\alpha=k$.
When approaching the ballistic limit beyond this point, traveling waves are
allowed with speeds lower than $v$. The nature of these will be analyzed in the
following discussion.

Around the equilibrium $(1,0)$ the eigenvalues are:
\begin{equation}
\mu_\mp=\frac{-c(k+\alpha)\mp\sqrt{(k+\alpha)^2c^2+4k\alpha m}}{2m}.
\end{equation}
Oscillations around this equilibrium are in principle possible, as observed in
\cite{MHK}, since there is no impediment to the solution growing beyond $U=1$.
The condition for oscillations is found from the radicand of the eigenvalues
$\mu_\mp$:
\begin{equation}
(k+\alpha)^2c^2+4k\alpha (v^2-c^2)< 0,
\end{equation}
which, after some manipulation, becomes
\begin{equation}
(\alpha-k)^2 c^2 +4k\alpha v^2 < 0.
\label{cosc}
\end{equation}
Both terms in this relation are positive, implying that the relation cannot be
satisfied by the wave speed for any set of parameters.

A similar analysis can be carried out with the equation~(\ref{mhkteleg}),
corresponding to MHK model. The character of the equilibria for both
generalizations is resumed in Tables~\ref{table1} and~\ref{table2}. Note that
there is a $c_{osc}$ in Table~\ref{table2} Case (7), not present in the results
resumed in Table~\ref{table1} because Eq.~(\ref{cosc}) cannot be satisfied.
Different combinations of parameters allow a variety of traveling fronts
connecting one equilibrium to the other. Cases (1) to (7) in Table~\ref{table2}
are the fully nonlinear equivalents of the piecewise linear situations analyzed
in \cite{MHK}. The situations resumed in Table~\ref{table1} are discussed
below.

\begin{table}[h]
\centering
\begin{tabular}{|c|c|c|c|c|}
\hline
Case No. & $\alpha, k$ & $c$ & $(0,0)$ & $(1,0)$ \\
\hline
\hline
(1) & & $c<c_{min}$ & stable spiral & saddle\\
\cline{1-1}\cline{3-5}
(2) & $\alpha>k$ & $c_{min}<c<v$ & stable node & saddle\\
\cline{1-1}\cline{3-5}
(3) &  & $v<c$ & saddle & unstable node\\
\hline
\hline
(4) &  & $c<c_{min}$ & unstable spiral & saddle\\
\cline{1-1}\cline{3-5}
(5) & $\alpha<k$ & $c_{min}<c<v$ & unstable node & saddle\\
\cline{1-1}\cline{3-5}
(6) &  & $v<c$ & saddle & unstable node\\
\hline
\end{tabular}
\caption{Character of the equilibria for the possible combinations of
parameters and wave velocity, according to the generalization studied in this
paper. The ``case number'' refers to the discussion in the text.}
\label{table1}
\end{table}

\begin{table}[h]
\centering
\begin{tabular}{|c|c|c|c|c|}
\hline
Case No. & $\alpha, k$ & $c$ & $(0,0)$ & $(1,0)$ \\
\hline
\hline
(1) & & $c<c_{min}$ & stable spiral & saddle\\
\cline{1-1}\cline{3-5}
(2) &$\alpha>k$ & $c_{min}<c<v$ & stable node & saddle\\
\cline{1-1}\cline{3-5}
(3) & & $v<c$ & saddle & unstable node\\
\hline
\hline
(4) & & $c<c_{min}$ & stable spiral & saddle\\
\cline{1-1}\cline{3-5}
(5) &$\alpha<k$ & $c_{min}<c<v$ & stable node & saddle\\
\cline{1-1}\cline{3-5}
(6) & & $v<c<c_{osc}$ & saddle & unstable node\\
\cline{1-1}\cline{3-5}
(7) & & $c_{osc}<c$ & saddle & unstable spiral\\
\hline
\end{tabular}
\caption{Character of the equilibria for the possible combinations of
parameters and wave velocity, according to the generalization studied in
\cite{MHK}.}
\label{table2}
\end{table}

Equation.~(\ref{oscillator2}) can be exploited directly to understand the
nature of the solutions of the traveling wave problem, without obtaining
analytical solutions via simplifications. We will base the following discussion
on the mechanical interpretation of Eq.~(\ref{oscillator2}), which can be
reinterpreted, following MHK, as describing the motion of a particle of mass
$m$ in a nonlinear potential, subject to a state-dependent damping that can be
positive or negative. Since the mass of the particle can also be positive or
negative, depending on whether the velocity $c$ is lower than $v$ or not, we
analyze the two cases separately.

\subsection{$c<v$}

In this case a particle of mass $m>0$ is moving in a potential $\phi(U)=\alpha
k (U^2/2-U^3/3)$ [see Fig.~\ref{potentials} (a)]. This potential has a minimum
at $U=0$, but we have ruled out solutions that oscillate around it based on the
positivity of $U$ [this are cases (1) and (4) in Table~\ref{table1}]. Another
possible solution is an overdamped trajectory that connects the equilibrium at
$U=1$ with that at $U=0$. This corresponds to a traveling front of the state 1
invading the system at state 0. The damping coefficient is
$\gamma=c(\alpha-k+2kU)$. We can see that if $\alpha>k$, then $\gamma>0$ for
all values of $U$ along this trajectory. This is a solution that connects the
saddle with the stable node shown in the case (2) of Table~\ref{table1}.

Now, if $\alpha<k$, the motion is damped on part of the trajectory [when
$U>(1-\alpha/k)/2$, see Fig.~\ref{potentials} (c)] and anti-damped on the rest,
specifically near $U=0$. This means that the equilibrium at $U=0$ is unstable,
and there is a solution connecting it to the equilibrium at $U=1$. It is a
trajectory connecting the unstable node at $(0,0)$ to the stable manifold of
the saddle at $(1,0)$ [see Table~\ref{table1}, case (5)]. This trajectory
corresponds to a front of state 0 invading the state 1.

\subsection{$v<c$}

\begin{figure}
\centering
\resizebox{\columnwidth}{!}{\includegraphics[bb=8 200 784 793]{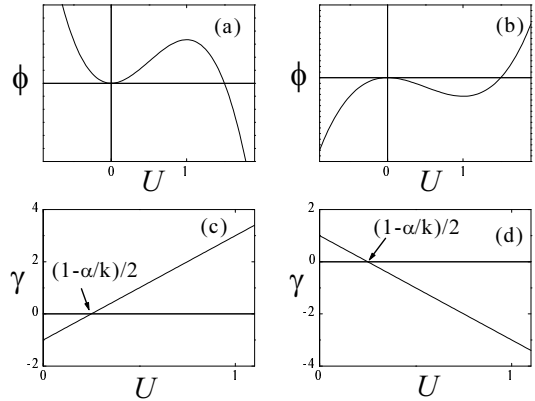}}
\caption{The potentials and the damping coefficients of the mechanical analog of
Eq.~(\ref{oscillator2}). Cases (a) and (c) correspond to $m>0$, while (b) and
(d) correspond to $m<0$.}
\label{potentials}
\end{figure}

In this situation $m$ is negative, so we multiply Eq.~(\ref{oscillator2}) by
$-1$, to obtain:
\begin{equation}
|m|U''-c(\alpha-k+2kU)U'=\alpha k U(1-U).
\end{equation}
Now the particle of mass $|m|$ moves in the potential $\phi(U)=-\alpha k
(U^2/2-U^3/3)$ [refer to Fig.~\ref{potentials} (b)], where there is a stable
equilibrium at $U=1$. The sign of the damping coefficient is also reversed, and
the motion is always anti-damped when $\alpha>k$. When $\alpha<k$, it is always
anti-damped in the vicinity of $U=1$. It can be seen in Table~\ref{table1}
[cases (3) and (6)] that the equilibrium at $(0,0)$ is always a saddle, and the
equilibrium at $(1,0)$ is always an unstable node. There is a trajectory that
connects the unstable node to the stable manifold of the saddle, corresponding
to a front of state 1 invading the state 0.

We note that, contrary to the generalized telegrapher's equation studied in
\cite{MHK}, we do not have here oscillating wave shapes (corresponding to the
unstable spiral at $(1,0)$ shown in Table~\ref{table2}). Also to be remarked
are the solutions where the state $U=0$---which is an unstable state of the
logistic equation--- invades the state $U=1$. This stabilization of the null
state has been made possible by interplay of memory and reaction, through the
state-dependent damping coefficient of Eq.~(\ref{telegraph}).

\begin{figure}
\centering
\resizebox{\columnwidth}{!}{\includegraphics[bb=8 200 784 793]{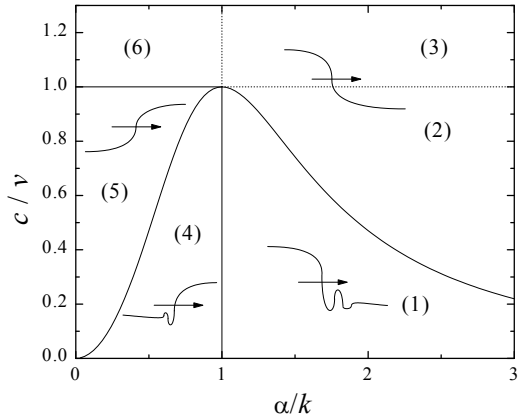}}
\caption{The regions in parameter space where different traveling fronts are found.
The front shapes are schematic. The numbers refer to the cases shown in
Table~\protect\ref{table1}.}
\label{phase}
\end{figure}

A phase diagram summarizing these results is shown in Fig.~\ref{phase}. A
qualitative indication of the nature of the front wave is given, together with
a reference to the cases enumerated in Table~\ref{table1}.

\section{Piecewise linearization}
\label{piecewise}

Explicit solutions corresponding to the front waves described above cannot be
found in the fully nonlinear situation. However, a piecewise linearization of
Eq.~(\ref{telegraph}) can be made, just as in \cite{MHK}, taking the following
as a reaction function:
\begin{equation}
f(U)=\left\{\begin{array}{ll}
            \displaystyle U/a,&\;\;\; U\leq a, \\
            \displaystyle (b-U)/(b-a),&\;\;\; U\geq a,
            \end{array}
     \right.
\label{flinear}
\end{equation}
where $a<b$. With this reaction term, which incidentally, generalizes somewhat
the logistic form (which corresponds to $a=b$), the oscillator equations in the
traveling waves ansatz (\ref{oscillator}) become:
\begin{eqnarray}
m\,U'' + 2\gamma_1\,U'+k_1^2\,U &=& 0,\;\;\;\;U\le a, \label{linosc1}\\
m\,U'' +2\gamma_2\,U'+k_2^2\,(b-U) &=& 0,\;\;\;\;U\ge a.
\label{linosc2}
\end{eqnarray}
where
\begin{equation}
\gamma_1=\frac{c}{2}\left(\alpha-\frac{k}{a}\right),\;\;\;
\gamma_2=\frac{c}{2}\left(\alpha-\frac{k}{b-a}\right),
\end{equation}
\begin{equation}
k_1^2=\frac{\alpha k}{a},\;\;\;
k_2^2=\frac{\alpha k}{b-a}.
\end{equation}

\begin{figure}
\centering
\resizebox{\columnwidth}{!}{\includegraphics[bb=8 200 784 793]{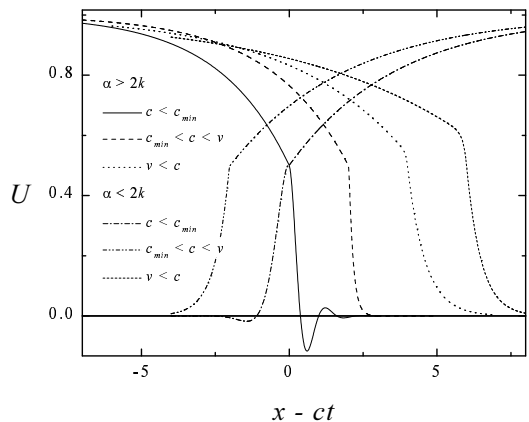}}
\caption{Front waves in the piecewise linear version of the model. $\alpha<2k$
and $\alpha>2k$ correspond to the two distinct regimes of damping or
anti-damping, in this linearized model. We have taken $a=1/2$ and $b=1$.}
\label{fronts}
\end{figure}

Suppose that we are looking for a front solution that interpolates from $U=b$
as $z\rightarrow -\infty$ to $U=0$ as $z\rightarrow\infty$. Without loss of
generality, we take $U=a$ at $z=0$. We look for solutions of two oscillators:
one to the right of $z=0$, and one to the left of $z=0$. We define the
following variables:
\begin{equation}
\begin{array}{lcll}
U_R(z)&=&U(z),&\;\;\;z\ge 0, \\
U_L(-z)&=&b-U(z),&\;\;\;z\ge 0. \\
\end{array}
\end{equation}
Equations~(\ref{linosc1}) and (\ref{linosc2}) become:
\begin{eqnarray}
m\,U_R'' + 2\gamma_1\,U_R' + k_1^2\,U_R &=& 0, \label{ur}\\
-m\,U_L'' + 2\gamma_2\,U_L' + k_1^2\,U_L &=& 0.
\label{ul}
\end{eqnarray}
The wave front shape, which is a solution to Eqs.~(\ref{linosc1}) and
(\ref{linosc2}), can be easily found in terms of exponentials, by matching
solutions of Eqs.~(\ref{ur}) and (\ref{ul}) and their derivatives on either
side of $z=0$. In some cases, one of the solutions of Eqs.~(\ref{ur}) or
(\ref{ul}) contains a growing exponential, and the corresponding factor has to
be set to zero to avoid unbounded growth. Examples of all the cases described
in the previous section are shown in Fig.~\ref{fronts}, within this piecewise
linear scheme. The solutions have been arbitrarily shifted in the variable
$z=x-ct$ to avoid superposition of the curves. In the model linearized
according to (\ref{flinear}), the regimes of damping and anti-damping of
Eq.~(\ref{oscillator}) are separated by the condition $\alpha=2k$, which is
used in the figure caption to classify the curves. (The condition separating
the damping from the anti-damping regimes is $\alpha=k/a$ in the piecewise
linear model, instead of the $\alpha=k$ of the fully nonlinear model.) The case
numbers corresponding to Table~\ref{table1} are also shown in the caption.

\section{Conclusions}
\label{conclusion}

Reaction-diffusion systems in which the transport process is wave-like at short
times and diffusive at long times are the focus of the present investigation.
This passage of the character of the motion from coherent to incoherent is a
general feature of all physical systems and may be represented by a memory
function whose decay time (or correlation time) represents the demarcation. We
have analyzed here a generalization of Fisher's reaction-diffusion equation
which has a logistic nonlinearity describing the reaction process. Our study
has centered on traveling wave solutions. From arguments without approximation
we have found a generalization of the known Fisher's equation result regarding
the minimum speed of the traveling waves. While the generalization converges to
the Fisher's equation result in the limit of diffusive transport (memory that
decays infinitely fast), sharp differences occur in the wave-limit (see
Fig.~\ref{cmin-fig}): new kinds of solutions are found to be possible involving
``inverse'' fronts, in which the state $U=0$ invades the state $U=1$ (see
Fig.~\ref{phase}). We have also found explicit analytic solutions (typical
cases plotted in Fig.~\ref{fronts}) via the piecewise linearization introduced
for this problem elsewhere~\cite{MHK} and pointed out a number of differences
in the predictions arising in our present analysis. Some of these differences
are expected but at least one of them is surprising: solutions which oscillate
spatially when they have a speed above a certain value (called $c_{osc}$ in
Ref.~\cite{MHK}; see also Table~\ref{table2}) predicted in Ref.~\cite{MHK} are
found to disappear in the present analysis which can be argued to be a more
natural generalization of Fisher's equation to include wave-like transport.
There are a number of issues, such as the stability of the solutions against
perturbations introduced into the system, which will be studied in future
investigations.

We mention in passing the problem of what kind of initial conditions eventually
evolve into a traveling front. The general problem can be very difficult, but a
simplified treatment is possible as follows. Consider that the initial state
has the behavior $u(x,0)\sim Ae^{-ax}$ for $x\rightarrow\infty$.
Correspondingly, we suppose that the leading edge of the traveling wave has the
form:
\begin{equation}
u(x,t)=A e^{-a(x-ct)},
\end{equation}
with $a$ and $A$ arbitrary.

We substitute this into the differential equation, and suppose that, at the
leading edge, $u\approx 0$:
\begin{equation}
(ac)^2\,u + \alpha ac\,u - kac\,u + o(u^2) = v^2a^2\,u + \alpha k \,u + o(u^2).
\end{equation}
Disregarding the terms in $u^2$ we arrive at a dispersion relation between $c$
and $a$:
\begin{equation}
c=\frac{k-\alpha}{2a}\pm\sqrt{\frac{(\alpha+k)^2}{4a^2}+v^2}.
\end{equation}

The negative sign before the square root in this expression leads to negative
$c$, and can be disregarded. The expression with the positive sign gives the
speed of the front wave that will eventually develop from an initial condition
that has an exponential decay of coefficient $a$.

\begin{acknowledgments}
This work was supported in part by the Los Alamos National Laboratory via a
grant made to the University of New Mexico (Consortium of the Americas for
Interdisciplinary Science) and by the National Science Foundation's Division of
Materials Research via grant No. DMR0097204. G. A. thanks the support of the
Consortium of the Americas for Interdisciplinary Science and the hospitality of
the University of New Mexico.
\end{acknowledgments}

\end{document}